\documentclass[aps,twocolumn,amsfonts,showpacs,pra,amsmath,amssymb]{revtex4-1}
\usepackage{graphicx}
\usepackage{dcolumn}
\usepackage{bm}
\usepackage[usenames]{color}
\usepackage{mathtools}
\usepackage{newtxtext}
\usepackage[varg]{newtxmath}
\usepackage[normalem]{ulem}
\usepackage{lipsum}
\begin{document}

\preprint{APS/123-QED}

\title{Optical Response in Excitonic Insulating State: Variational Cluster Approach }


\newcommand{\TohokuUniv}{Department of Physics, Tohoku University, Sendai 980-8578, Japan}
\newcommand{\Waseda}{Waseda Institute for Advanced Study, Waseda University, Tokyo 169-8050, Japan}
\newcommand{\TohokuUni}{Department of Physics, Tohoku University, Sendai 980-8578, Japan}

\author{Hengyue Li}
\affiliation{\TohokuUniv}

\author{Makoto Naka}
\affiliation{\Waseda}

\author{Junya Otsuki}
\altaffiliation[Present address: ]{Research Institute for Interdisciplinary Science, Okayama University, Okayama 700-8530, Japan}
\affiliation{\TohokuUniv}

\author{Sumio Ishihara}
\affiliation{\TohokuUni}


\date{\today}

\begin{abstract}


Optical responses in an excitonic insulating (EI) system with strong electron correlation 
are studied. 
We adopt the two-orbital Hubbard model with a finite energy difference between the two orbitals 
where the spin state degree of freedom exists. 
This model is analyzed by the variational cluster approach.  
In order to include the local electron correlation effect, 
the vertex correction is taken into account in the formulation of the optical conductivity spectra. 
We calculate a finite-temperature phase diagram, in which an EI phase appears 
between a low-spin band insulating state and a high-spin Mott insulating state. 
Characteristic components of the optical conductivity spectra consisting of a sharp peak and continuum appear in the EI  phase. 
Integrated intensity almost follows the order parameter of the EI state, suggesting that this component is available to identify the EI phases and transitions.


\end{abstract}

\maketitle


\section{Introduction }

Excitonic insulating (EI) state was proposed more than a half-century ago in semiconductors and semimetals and has been 
studied intensively in both the experimental and theoretical sides \citep{mott1961,zittartz1967theory,PhysRev.158.462,RevModPhys.40.755}. 
When the attractive Coulomb interaction between electrons and holes
overcomes the insulating gap energy, 
the electron-hole pairs are produced spontaneously, and a macroscopic number of the excitions are condensed in low temperatures.    
Some conceptual similarities of this excitonic condensation to the superconductivity and the charge-density wave 
have been examined. 
Owing to the recent great progresses of the experimental technique, the study of the EI state is revived. 
Prototypical examples are 1T-TiSe$_{2}$ \citep{EI_1TTise2}
and TaNiSe$_{5}$ \citep{EITa2NiSe5}, 
in which flattenings of the top of the valence band dispersions were observed below the structural-phase transition temperatures by the angular-resolved photoemission spectroscopy (ARPES). 

Another candidate material for the EI state is the cobalt oxides with the perovskite crystal structure, $R_{1-x}A_{x}$CoO$_{3}$, where $R$ and $A$ are a rare-earth ion and an alkaline earth ion, respectively.
This series of materials is widely recognized to show the spin cross over phenomena by changing temperature, pressure and magnetic field~\citep{tokura1998, asai1998two}. 
In a single Co$^{3+}$,  the three kinds of electron configurations are possible: $(t_{{\rm 2g}})^{6}(e_{{\rm g}})^{0}$ with $S=0$, $(t_{{\rm 2g}})^{5}(e_{{\rm g}})^{1}$ with $S=1$ and
$(t_{{\rm 2g}})^{4}(e_{{\rm g}})^{2}$ with $S=2$, which are termed
the low-spin (LS), intermediate spin and high-spin (HS) states, respectively. 
Stability of the three spin states is governed dominantly by the competition of the crystalline field splitting and the Hund coupling. 
Recently, an EI state is proposed in Pr$_{1-x}$Ca$_{x}$CoO$_{3}$, where a metal-insulator transition was observed  around $90$K by the magnetization, the specific heat and the
structural analyses~\citep{PhysRevB.66.052418,PhysRevB.69.144406}.  
Because of no evidence of the charge and/or orbital orders so far, an EI state is considered to be a possible origin of this  transition. 
An EI state is also proposed in LaCoO$_{3}$ under a high magnetic field around 100T~\citep{ikeda2016}. 
A number of the theoretical examinations were performed to clarify the EI state in the perovskite cobaltites, and a relation to the spin-state degree of freedom in the Co ions~\citep{PhysRevB.89.115134, PhysRevB.90.235112, EInasu, tatsuno2016}. 

One of the unresolved issues in the EI researches is to develop the experimental way to identify the
EI state and transition. 
The flattening of the top of the valence-band dispersion observed by ARPES suggests the EI transition, 
although this is not the direct evidence of the EI transition. 
Photoinduced transient dynamics provide rich information for the driving force of the structural phase transition 
by utilizing different time scales of the electronic and lattice degrees of freedom~\citep{lu2017zero, mor2017, werdehausen2018}. 
Once the direct method to identify the EI phase is settled, 
this technique will be applied to a wide range of candidate materials. 

In this paper, we study the optical responses in EI states. The two-orbital
Hubbard model with a finite energy difference between the orbitals is
analyzed by utilizing the variational cluster approach. The optical
conductivity spectra are formulated in the two-particle Green's function,
in which the vertex corrections are taken into account. 
The finite temperature ($T$) phase diagram and the one-particle excitation spectra
are obtained in  the EI phase. 
A characteristic structure in the optical conductivity spectra emerges in
the EI state due to the hybridization between the two orbitals. 
The intensity of this structure in low temperatures almost follows the EI order parameter. 
We propose that this is available to identify the EI state. 

In Sec.~II, the adopted two-orbital Hubbard model with the finite energy
difference is introduced. The variational-cluster approach (VCA) applied to the model 
is introduced in Sec.~IIIA, and the formulation of the optical conductivity spectra 
with the vertex correction is explained in Sec.~IIIB. 
The numerical results of the finite $T$ phase diagram, and the excitation spectra are presented in Sec.~IV.
Section V is devoted to concluding remarks.

\section{Model}

\begin{figure}[t]
\begin{center}
\includegraphics[width=0.5\columnwidth,clip]{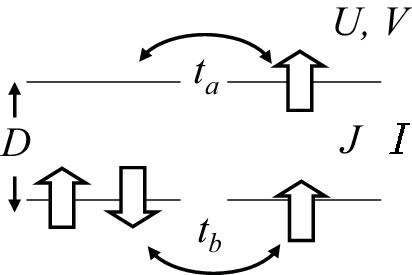}
\end{center}
\caption{(Color online) 
A schematic illustration of the Hamiltonian in Eq. ~(\ref{eq:2-4}). 
Horizontal bars and bold arrows represent the two orbitals, and the electron spins, respectively.}
\label{H} 
\end{figure}

We start from the two-orbital Hubbard model with the finite energy difference between the two orbitals 
given as 
\begin{align}
H =H_{L}+H_{t} ,
\label{eq:2-4}
\end{align}
where $H_{L}$ represents the local part and $H_{t}$ represents the electron hoppings between different sites. 
The first term is given as 
\begin{align}
H_{L}=\sum_{i} \left [h_{0}(i)+h_{\textrm{int}}(i) \right ] ,
\end{align}
with 
\begin{align}
h_{0}(i) & =\frac{D}{2}\sum_{\sigma}(n_{ia\sigma}-n_{ib\sigma})-\mu\sum_{\alpha\sigma}n_{i\alpha\sigma},\\
h_{\textrm{int}}(i) & =U\sum_{\alpha}n_{i\alpha\uparrow}n_{i\alpha\downarrow}+Vn_{ia}n_{ib}\nonumber \\
 & +J\sum_{\sigma\sigma'}c_{ia\sigma}^{\dagger}c_{ib\sigma'}^{\dagger}c_{ia\sigma'}c_{ib\sigma}+I\sum_{\alpha}c_{i\alpha\uparrow}^{\dagger}c_{i\alpha\downarrow}^{\dagger}c_{i\bar{\alpha}\downarrow}c_{i\bar{\alpha}\uparrow}.\label{eq:2model-5}
\end{align}
We introduce  the creation (annihilation) operator $c_{i\alpha\sigma}^\dagger$ ($c_{i\alpha\sigma}$) of an electron
with spin $\sigma$($=\uparrow,\downarrow$) and orbital $\alpha$($=a,b$)
at site $i$, the number operator $n_{i \alpha}=\sum_{\sigma}n_{i\alpha\sigma}=\sum_{\sigma}c_{i\alpha\sigma}^{\dagger}c_{i\alpha\sigma}$, 
and $\bar \alpha=a (b)$ for $\alpha=b (a)$. 
We define that $D$ is energy difference between orbitals $a$ and $b$, 
$U$ and $V$ are the intra- and inter-orbital Coulomb interactions, respectively, 
$J$ is the Hund's coupling, $I$ is the pair hopping, 
and $\mu$ is the chemical potential. 
The second term in Eq. (\ref{eq:2-4}) is given as 
\begin{align}
H_{t}=\sum_{\langle i j \rangle }h_{t}(i,j),
\end{align}
with 
\begin{align}
h_{t}(i,j)  =
\sum_{\alpha\sigma}t_{\alpha}c_{i\alpha\sigma}^{\dagger}c_{j\alpha\sigma} , 
\end{align}
for the nearest-neighbor (NN) sites, 
where $t_{\alpha}$ represents the hopping integral between the orbital $\alpha$. 
The hopping integrals are assumed to be diagonal with respect to the orbital. 
The present Hamiltonian is schematically shown in Fig.~\ref{H}.

In the numerical calculations, 
we consider a two-dimensional square lattice with the number of sites $N$. 
The two hopping integrals are assumed to be equal with each other 
and are set to the energy unit as 
$t \equiv t_{a}=t_{b}=1$. 
Other parameters are set to be $D=9$, $U=5J$, $V=3J$, and $I=J$, 
where $J$ is varied. 
The electron number is fixed to be half filling, 
realized by choosing the chemical potential at $\mu=(U+2V-J)/2$, 
which satisfies the particle-hole symmetry in the system. 

\section{Method}

In order to examine electronic structures in various phases emerging due to local
electronic correlations, we employ VCA developed by Potthoff \textit{et al}. \citep{VCA2003}. 
In this section, we present an introduction of VCA and its
application to the present model. 
We also present the method to calculate the optical conductivity  where the vertex correction is taken into account.

\subsection{A brief introduction of VCA}

The VCA is a general framework of cluster theories, which includes
the cluster perturbation theory (CPT) \citep{VCA2003,CPT2000} and
the cluster extension of the dynamical mean-field theory (DMFT) \citep{DMFT,PhysRevB.61.12739,PhysRevLett.87.186401}
depending on the number of uncorrelated ``bath'' sites (zero for
the former and infinity for the latter) \citep{VCA2003}. In the case
without bath sites, the VCA forms a generalization of the CPT where
we can introduce local potentials as a variational parameter, which
allows us to address symmetry broken phases. This special case is
called variational CPT (V-CPT) \citep{VCA_2004_AF}. The VCA, in particular,
V-CPT has been successfully applied to many systems such as the antiferromagnetic (AF) phase
\citep{VCA_2004_AF}, charge ordering \citep{VCA_2004_CO}, high-$T_{c}$
cuprates \citep{VCA2005}, spin liquid \citep{VCA_2008_spinliquid},
correlated topological insulators \citep{VCA_2011_toplogicalinsulator},
bosonic systems and Bose-Einstein condensation \citep{VCA_2011_boson,VCA_2011_boson2,VCA_2006_bosonic}
and non-equilibrium system \citep{VCA_2011_nonequ1,VCA_2013_nonequ2}.
Detailed explanations of the method can be found in Ref. \citep{VCA2003,VCA2003_1,VCA2003_2,VCA_2004_AF,VCA_2004_CO,VCA2008,QCM2008}.
In the following, we give a brief summary of the method.

The VCA is formulated based on the self-energy functional theory (SFT)
\citep{VCA2003_1,VCA2003_2}. In the SFT, we consider the self-energy
functional of the thermodynamic potential, $\Omega[\boldsymbol{\Sigma}]$,
which is given by 
\begin{align}
\Omega[\boldsymbol{\Sigma}] & =\mathcal{F}_{U}[\boldsymbol{\Sigma}]-\sum_{\boldsymbol{k}}\text{Tr}\ln\{-[\boldsymbol{\mathcal{G}}_{0}^{-1}(\boldsymbol{k})-\boldsymbol{\Sigma}]^{-1}\},\label{eq:4}
\end{align}
where $\mathcal{F}_{U}[\boldsymbol{\Sigma}]$ is the Legendre transform
of the Luttinger-Ward functional $\Phi[\boldsymbol{\mathcal{G}}]$
\citep{LuttingerFunction}, $\mathcal{\boldsymbol{G}}_{0}$ ($\boldsymbol{\mathcal{G}}$)
is the non-interacting (interacting) Green's function, $\boldsymbol{\Sigma}$
is the trial self-energy, and $\boldsymbol{k}$ is the wave vector
in the Brillouin zone. The bold symbol represents matrices with the spin
and orbital indices, and Tr stands for summations over these indices
and the Matsubara frequencies: $\text{Tr}\equiv T\sum_{\omega_{n}}\sum_{\alpha\sigma}$.
The subscript $U$ in $\mathcal{F}_{U}$ indicates that $\mathcal{F}_{U}$
depends only on the two-body (interaction) terms, but not on the one-body
terms. According to the SFT \citep{VCA2003_1,VCA2003_2}, the functional
$\Omega[\boldsymbol{\Sigma}]$ holds the variational principle with
respect to $\boldsymbol{\Sigma}$, $\delta\Omega[\boldsymbol{\Sigma}]/\delta\boldsymbol{\Sigma}=0$,
as in the Luttinger-Ward formalism in terms of $\boldsymbol{\mathcal{G}}$
\citep{LuttingerFunction}. In other words, our task is to find the
stationary point of Eq. (\ref{eq:4}) rather than to solve the original
model directly.

\begin{figure}\centering
\includegraphics[scale=0.35]{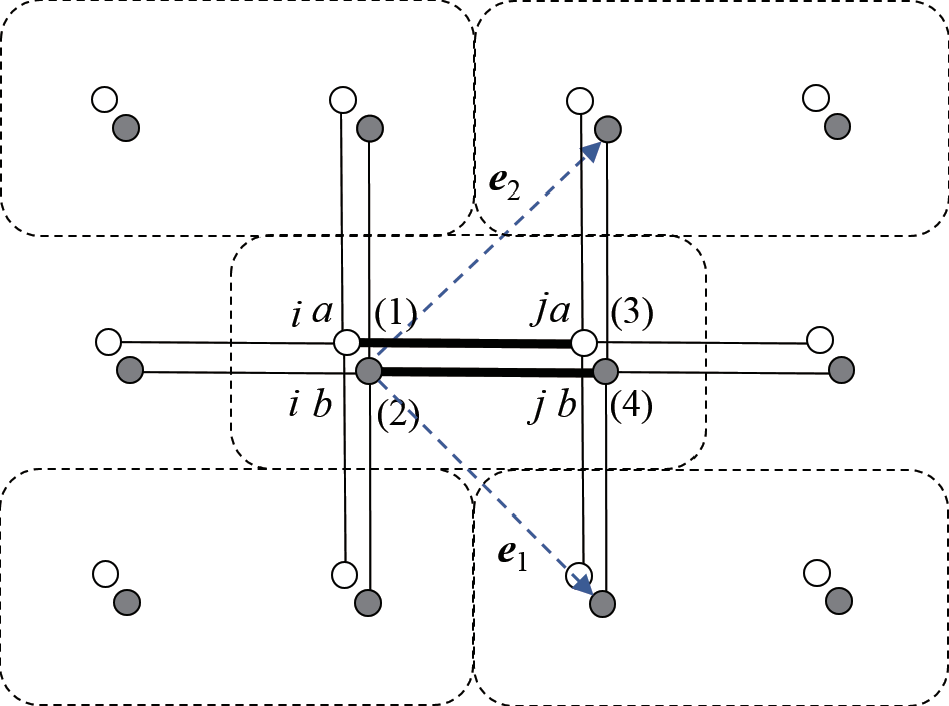}
\caption{The present reference system. 
Broken squares represent unit cells. 
Open and filled circles represent the orbital $a$ and $b$, respectively. 
Labels $(1)$-$(4)$ indicate the independent orbitals and sites in a cluster.}
\label{Hr} 
\end{figure}

The functional $\mathcal{F}_{U}[\boldsymbol{\Sigma}]$ takes account
of interactions, but is essentially impossible to be evaluated exactly
for arbitrary form of $\boldsymbol{\Sigma}$. In VCA, we instead evaluate
$\mathcal{F}_{U}[\boldsymbol{\Sigma}]$ using an auxiliary cluster
model called ``reference'' system. The Hamiltonian $H'$ of the
reference system is given by the form 
\begin{align}
H' & =H_{L}+H'_{t},\label{eq:H'}
\end{align}
where $H'_{t}$ contains arbitrary one-body terms (hopping and local
potential). In particular, we can consider a situation where some
hopping integrals are absent so that the lattice is divided into clusters which
are arranged periodically (see Fig.~\ref{Hr}). When the cluster is
small enough, we can diagonalize $H'$ and evaluate physical quantities
such as the thermodynamic potential $\Omega'$ 
and the Green's function $\boldsymbol{G}'$ in the reference system.
Furthermore, $H'_{t}$ can include any potentials that break the symmetries 
of the original Hamiltonian, leading to a description of the symmetry
broken states.

The important point in Eq. (\ref{eq:H'}) is that the interaction
$H_{L}$ is the same as in $H$. It leads that the functional $\mathcal{F}_{U}[\boldsymbol{\Sigma}]$
is common between the original and reference system. Therefore, we
can eliminate $\mathcal{F}_{U}[\boldsymbol{\Sigma}]$ from the equations
for $\Omega$ and $\Omega'$, and obtain 
\begin{align}
\Omega[\boldsymbol{\Sigma}] & =\Omega'[\boldsymbol{\Sigma}]-\sum_{\boldsymbol{q}}\text{Tr}\ln(\boldsymbol{1}-\boldsymbol{V}_{\boldsymbol{q}}\boldsymbol{G}'),\label{eq:5}
\end{align}
where $\boldsymbol{q}$ is the reduced wave vector, when the cluster
contains more than one site, and $\boldsymbol{V}_{\boldsymbol{q}}=\boldsymbol{G}_{0}^{'-1}-\boldsymbol{\mathcal{G}}_{0}^{-1}(\boldsymbol{q})$
is the difference of the signal-particle terms between the original
system and the reference system.

All quantities on the right-hand side of Eq.~(\ref{eq:5}) can be
computed numerically with use of the reference system, meaning that
we can evaluate $\Omega[\boldsymbol{\Sigma}]$ \emph{exactly}. However,
the argument $\boldsymbol{\Sigma}$ cannot be changed arbitrary, but
is determined only through one-body parameters in $H'_{t}$, namely,
$\boldsymbol{\Sigma}=\boldsymbol{\Sigma}(t')$, where $t'$ indicates
a parameter in $H'_{t}$. We will search for the stationary point
of $\Omega[\boldsymbol{\Sigma}]$ within this \emph{restrected} space
of $\boldsymbol{\Sigma}(t')$ by $\partial\Omega[\boldsymbol{\Sigma}(t')]/\partial t'=0$.



\subsection{Reference system}
\label{sec:reference_system}

As the reference system, we make a minimal choice that is required
for discussing staggered ordered states. We retain hopping only between
a pair of sites so that two-site clusters are isolated as shown in
Fig.~\ref{Hr}. The one-body part $H'_{t}$ in the Hamiltonian, Eq.~(\ref{eq:H'}),
now reads 
\begin{align}
H'_{t}=\sum_{R}\left[\sum_{i,j\in R}h_{t}(i,j)+\sum_{i\in R}h_{\Delta}(i)\right],\label{eq:H'2}
\end{align}
where $\boldsymbol{R}$ is the index which specifies clusters. The
second term, $h_{\Delta}(i)$, is introduced to represent symmetry-breaking
local potentials, which will be discussed in Sec.~\ref{sec:order_parameters}.

It is convenient to introduce a combined site-orbital index, $(i, \alpha)\equiv l=1,...,4(\equiv L)$,
in each cluster (see Fig.~\ref{Hr}), and define an $L$-dimensional
vector $\psi_{\boldsymbol{R}\sigma}$ of annihilation or creation
operators such as 
$\psi_{\boldsymbol{R}\sigma}=(c_{1\sigma},...,c_{L\sigma})^{\textrm{T}}$.
Using this representation, $H'$ in Eq.~(\ref{eq:H'}) is represented
as 
\begin{align}
H' & =\sum_{\boldsymbol{R}\sigma}\psi_{\boldsymbol{R}\sigma}^{\dagger}\left(\boldsymbol{\mathcal{H}}+\boldsymbol{\Delta}^{\sigma}\right)\psi_{\boldsymbol{R}\sigma}+\sum_{i}h_{\textrm{int}}(i),\label{eq:10}
\end{align}
where $\boldsymbol{\mathcal{H}}$ and $\boldsymbol{\Delta}^{\sigma}$
denote the matrix elements of the one-body part, $h_{t}(i,j)+h_{L0}(i)$
and $h_{\Delta}(i)$, respectively. The explicit expression of $\boldsymbol{\mathcal{H}}$
is given by 
\begin{align}
\boldsymbol{\mathcal{H}} & =\left(\begin{array}{cccc}
\frac{D}{2}-\mu & 0 & t_{a} & 0\\
0 & -\frac{D}{2}-\mu & 0 & t_{b}\\
t_{a} & 0 & \frac{D}{2}-\mu & 0\\
0 & t_{b} & 0 & -\frac{D}{2}-\mu
\end{array}\right).
\label{eq:mathcal_H}
\end{align}
Including $h_{\textrm{int}}$, we diagonalize $H'$ by means of the
exact diagonalization method, and compute $\boldsymbol{G}'(\omega)$.

In order to calculate $\Omega$ in Eq.~(\ref{eq:5}), we further need
the expression of $\boldsymbol{V}_{\boldsymbol{q}}=\boldsymbol{G}_{0}^{'-1}-\boldsymbol{\mathcal{G}}_{0}^{-1}(\boldsymbol{q})$,
which represents the difference of the non-interacting Hamiltonian
in the reference and original systems, i.e., $H-H'$. 
More precisely, its bilinear form, $H-H'=\sum_{\boldsymbol{R}\sigma}\psi_{\boldsymbol{R}\sigma}^{\dagger}\boldsymbol{V}_{\boldsymbol{q}}^{\sigma}\psi_{\boldsymbol{R}\sigma}$,
yields the representation for $\boldsymbol{V}_{\boldsymbol{q}}^{\sigma}$
(depends on $\sigma$ because of $\boldsymbol{\Delta}^{\sigma}$).
We thus obtain 
\begin{align}
\boldsymbol{V}_{\boldsymbol{q}}^{\sigma}=\boldsymbol{\mathcal{T}_{q}}-\boldsymbol{\Delta}^{\sigma},\label{eq:V_q}
\end{align}
where $\boldsymbol{\mathcal{T}_{q}}$ denotes hopping which connects
different clusters. 
Its explicit expression is given by 
\begin{align}
\boldsymbol{\mathcal{T}_{q}} & =\left(\begin{array}{cccc}
0 & 0 & t_{a}E_{\boldsymbol{q}} & 0\\
0 & 0 & 0 & t_{b}E_{\boldsymbol{q}}\\
t_{a}E_{\boldsymbol{q}}^{*} & 0 & 0 & 0\\
0 & t_{b}E_{\boldsymbol{q}}^{*} & 0 & 0
\end{array}\right).
\label{eq:mathcal_T}
\end{align}
where $E_{\boldsymbol{q}}=e^{i\boldsymbol{q}\cdot\boldsymbol{e}_{1}}+e^{i\boldsymbol{q}\cdot\boldsymbol{e}_{2}}+e^{i\boldsymbol{q}\cdot(\boldsymbol{e}_{1}+\boldsymbol{e}_{2})}$ with 
$\boldsymbol{e}_{1}$ and $\boldsymbol{e}_{2}$ being the bases of the
super-lattice as shown in Fig. \ref{Hr}.

\subsection{Order parameters}

\label{sec:order_parameters}

In the present study, we consider three kinds of long-range orders: the AF state \citep{VCA_2004_AF}, the EI 
state \citep{EI_VCA_2011, EIVCA2012}, and the HS-LS ordered (HL) state \citep{EInasu, tatsuno2016}.
Accordingly, we define three Weiss fields given by 
\begin{align}
H_{\Delta}\equiv\sum_{i}h_{\Delta}(i)=\sum_{O}\delta_{O}p_{O},\label{eq:H-Delta}
\end{align}
where $O$ is either AF, EI, or HL. The potential $\delta_{O}$ is
a real number to be determined by the variational principle, and the
operator $p_{O}$ is defined by 
\begin{align}
p_{\textrm{AF}} & =\sum_{i \alpha}(-1)^{i} \left(n_{i\alpha\uparrow}-n_{i\alpha\downarrow} \right),\\
p_{\textrm{EI}} & =\sum_{i \sigma} e^{i\boldsymbol{Q}\cdot\boldsymbol{r}_{i}}(-1)^\sigma \left (c_{ia\sigma}^{\dagger}c_{ib\sigma}+h.c. \right), \\
p_{\textrm{HL}} & =\sum_{i \alpha\sigma}e^{i\boldsymbol{Q}\cdot\boldsymbol{r}_{i}}(-1)^{\alpha}c_{i\alpha\sigma}^{\dagger}c_{i\alpha\sigma},
\end{align}
with $\boldsymbol{Q}=(\pi,\pi)$.
We note that $p_{\rm AF}$ and $p_{\rm EI}$ imply the staggered N$\rm \acute e$el order and the excitonic order, respectively, and $p_{\rm HL}$ represents the spatial alignment of the HS and LS states.

By transforming $H_{\Delta}$ into the $L$-dimensional vector representation,
$H_{\Delta}=\sum_{\boldsymbol{R}\sigma}\psi_{\boldsymbol{R}\sigma}^{\dagger}\boldsymbol{\Delta}^{\sigma}\psi_{\boldsymbol{R}\sigma}$,
we obtain the expression of $\boldsymbol{\Delta}^{\sigma}$ as 
\begin{align}
\boldsymbol{\Delta}^{\sigma}=\sum_{O}\delta_{O}\boldsymbol{\Delta}_{O}^{\sigma},
\end{align}
with 
\begin{align}
\boldsymbol{\Delta}_{\textrm{AF}}^{\sigma} & =\sigma\left(\begin{array}{cccc}
1 & 0 & 0 & 0\\
0 & 1 & 0 & 0\\
0 & 0 & -1 & 0\\
0 & 0 & 0 & -1
\end{array}\right),\\
\boldsymbol{\Delta}_{\textrm{EI}}^{\sigma} & =\sigma\left(\begin{array}{cccc}
0 & 1 & 0 & 0\\
0 & 0 & 0 & 0\\
1 & 0 & 0 & -1\\
0 & 0 & -1 & 0
\end{array}\right),\\
\boldsymbol{\Delta}_{\textrm{HL}}^{\sigma} & =\left(\begin{array}{cccc}
1 & 0 & 0 & 0\\
0 & -1 & 0 & 0\\
0 & 0 & -1 & 0\\
0 & 0 & 0 & 1
\end{array}\right).
\end{align}
We thus have all ingredients for evaluation of $\Omega$ in Eq.~(\ref{eq:5}).

\begin{figure}[t] \centering
\includegraphics[width=1\columnwidth]{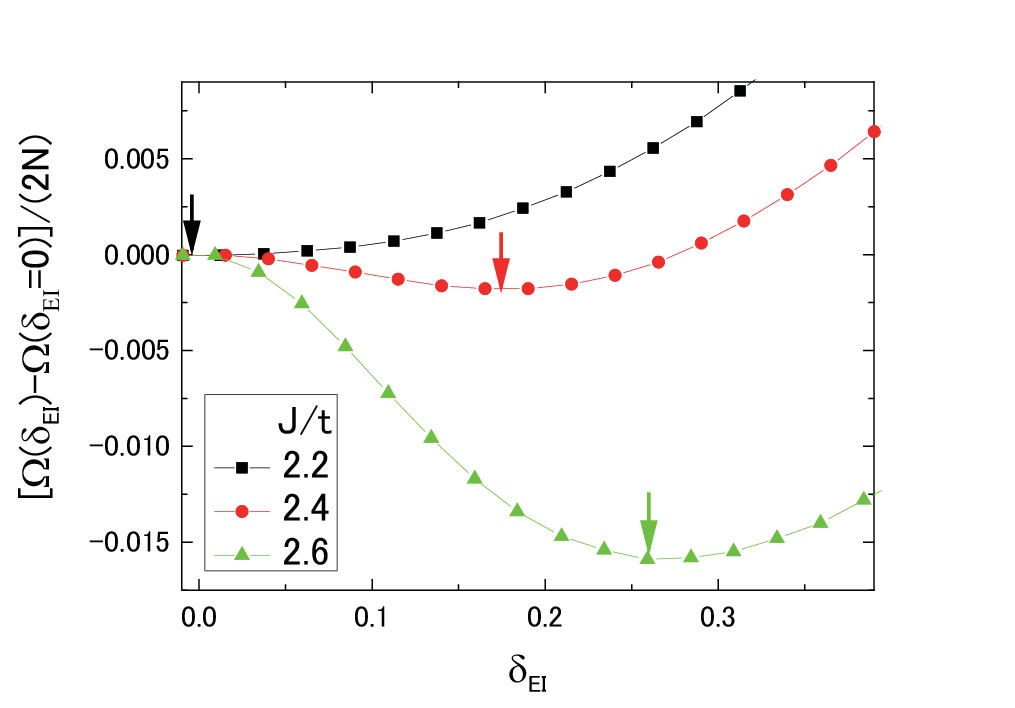}
\caption{(Color online) The ground potential $\Omega/2N$ as a function of the potential for the EI order $\delta_{\textrm{EI}}$
for various values of $J/t$. 
The arrows indicate the stationary points. 
We set $T/t=0.1$. 
}
\label{varexp} 
\end{figure}
\subsection{Computation procedure}

Having the expressions above, the variational principle is represented
as 
$\partial\Omega/\partial\delta_{O}=0$ for each field $\delta_{O}$.
We summarize below the standard steps for the VCA calculation: 
\begin{enumerate}
\item[(i)] For trial values of $\delta_{O}$, we solve the reference system whose
Hamiltonian $H'$ is given by Eq. (\ref{eq:10}), and compute $\boldsymbol{G}'(\omega)$
and $\Omega'$. 
\item[(ii)] We calculate $\Omega$ in the lattice system using Eqs.~(\ref{eq:5})
and (\ref{eq:V_q}). 
\item[(iii)] \label{item:find_stationary} We repeat the above steps by varying the
values of $\delta_{O}$ to find the stationary point of $\Omega$.
Thus, the physical self-energy $\boldsymbol{\Sigma}$ is obtained. 
\item[(iv)] At the stationary point, 
using the self-energy $\boldsymbol{\Sigma}$ in the reference system,
we calculate the lattice Green's function by 
$\mathcal{\boldsymbol{G}}(\boldsymbol{q},\omega)=[\boldsymbol{\mathcal{G}}_{0}^{-1}(\boldsymbol{q},\omega)-\boldsymbol{\Sigma}(\omega)]^{-1}$
and other physical quantities. 
\end{enumerate}

To illustrate step (iii), we show numerical results for $\Omega/2N$
as a function of $\delta_{\textrm{EI}}$ in Fig. \ref{varexp}. At
$J/t=2.2$, the stationary point is located at $\delta_{EI}=0$, meaning that
there is no symmetry breaking. For larger $J$, i.e., $J/t=2.4$ and
2.6, another stationary point appears around $\delta_{\textrm{EI}}\simeq0.175$
and $0.26$, respectively. In this way, we conclude that the EI phase
is realized for $J/t\gtrsim2.3$. More precise analysis will be presented
in Sec.~\ref{sec:Results}

\subsection{Optical conductivity with vertex correction}

\begin{figure*}
\includegraphics[scale=0.5]{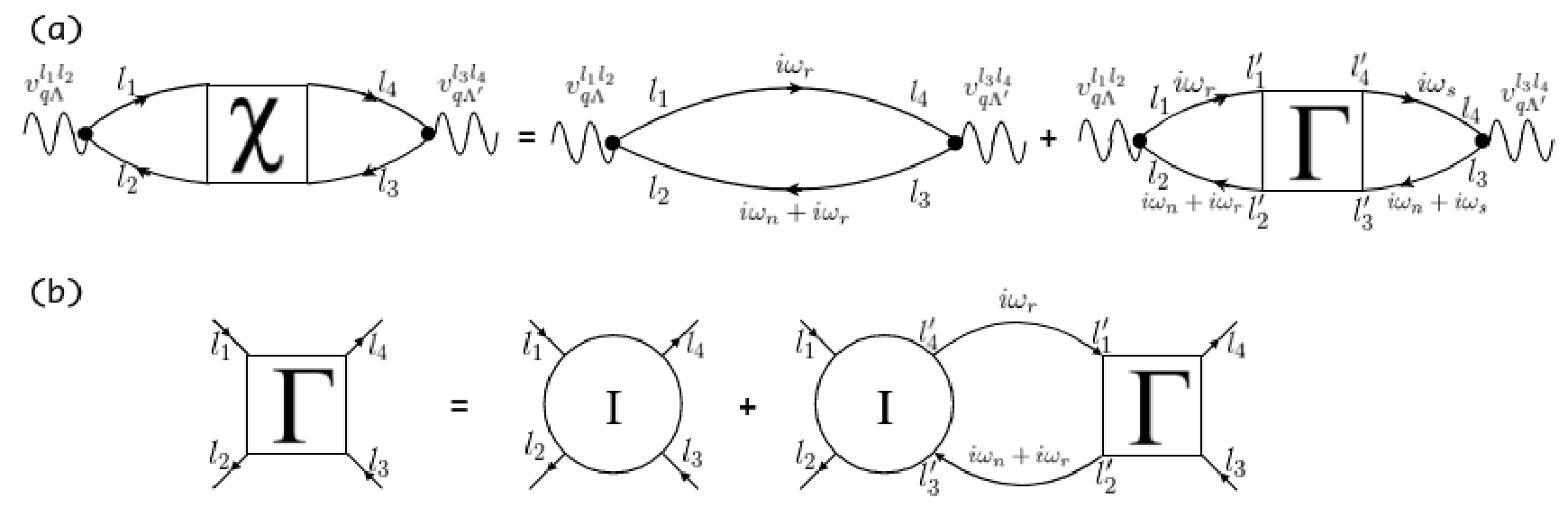}
\caption{(a) The Feynman diagrams of the two-particle Green's function, and 
(b)  those for the Bethe-Salpeter equation for the vertex function. 
A symbol $I$ represents the irreducible vertex function.}
\label{d1} 
\end{figure*}


We first introduce the current operator $j_{\Lambda}$ in the $\Lambda(=x, y, z)$ direction.
In the cluster representation, $j_{\Lambda}$ is given by 
\begin{align}
j_{\Lambda} & =i\sum_{\boldsymbol{q}\sigma}\sum_{ll'}c_{\boldsymbol{q}l\sigma}^{\dagger}v_{\boldsymbol{q}\Lambda}^{ll'}c_{\boldsymbol{q}l'\sigma} , 
\label{eq:13}
\end{align}
where $\boldsymbol{q}$ denotes the momentum in the reduced Brillouin zone,
and $l$ is the combined site-orbital index within the cluster (see Sec.~\ref{sec:reference_system}).
The velocity $v_{\boldsymbol{q}\Lambda}^{ll'}$ is given by
\begin{align} 
v_{\boldsymbol{q} \Lambda}^{ll'}=&-\sum_{\boldsymbol{R}}
e^{-i\boldsymbol{q}\cdot\boldsymbol{R}}
(\boldsymbol{\mathcal{T}}_{\boldsymbol{R}})_{ll'} [\boldsymbol{R}-(\boldsymbol{r}_{l}-\boldsymbol{r}_{l'})]_{\Lambda}
\nonumber \\
&+(\mathcal{H})_{ll'}(\boldsymbol{r}_{l}-\boldsymbol{r}_{l'})_{\Lambda},
\end{align}
where the matrices $\mathcal{H}$ and $\boldsymbol{\mathcal{T}}_{\boldsymbol{R}}$ are defined in Eq.~(\ref{eq:mathcal_H}) and Eq.~(\ref{eq:mathcal_T}), respectively.
The vector $\boldsymbol{r}_{l}$ represents the position of site $i$ corresponding to index $l\equiv (i,\alpha)$.
From the Kubo formula, the optical conductivity $\sigma_{\Lambda\Lambda'}(\omega)$ can evaluated by 
\begin{align}
\sigma_{\Lambda \Lambda'}(\omega) & =\frac{e^{2}}{N}\frac{1}{\omega}\text{Im}B_{\Lambda \Lambda'}(\omega+i\eta),
\end{align}
where $\eta$ is an infinitesimal small constant.
$B_{\Lambda\Lambda'}(\omega)$ is the current-current
correlation function defined by
\begin{align}
B_{\Lambda\Lambda'}(i\omega_n)=-\int_0^{\beta} d\tau \langle j_{\Lambda}(\tau)j_{\Lambda'}\rangle.
\end{align}
Replacing $j_{\Lambda}$ with Eq.~(\ref{eq:13}),
we obtain the explicit expression for $B_{\Lambda\Lambda'}(i\omega_n)$ as
\begin{align}
B_{\Lambda \Lambda'}(i\omega_n) & = \sum_{\boldsymbol{q}\boldsymbol{q}'}\sum_{l_1 l_2 l_3 l_4}\sum_{\sigma\sigma'}
v_{\boldsymbol{q}\Lambda}^{l_1 l_2}
\chi_{l_1 l_2 l_3 l_4}^{\boldsymbol{q}\sigma, \boldsymbol{q}'\sigma'}(i\omega_n)
v_{\boldsymbol{q}'\Lambda'}^{l_4 l_3},\label{eq:16}
\end{align}
where 
\begin{align}
\chi_{l_1 l_2 l_3 l_4}^{\boldsymbol{q}\sigma, \boldsymbol{q}'\sigma'}(i\omega_n)
&= \int_0^{\beta} d\tau \langle
c_{\boldsymbol{q} l_1 \sigma}^{\dagger}(\tau)
c_{\boldsymbol{q} l_2 \sigma}(\tau) 
c_{\boldsymbol{q}' l_4 \sigma'}^{\dagger}
c_{\boldsymbol{q}' l_3 \sigma'} \rangle
\nonumber \\
&\equiv [\boldsymbol{\chi}^{\boldsymbol{q}\boldsymbol{q}'}(i\omega_n)]_{(\sigma l_1 l_2), (\sigma' l_3 l_4)},
\end{align}
is the two-particle Green's function.
We introduced a matrix notation $\boldsymbol{\chi}^{\boldsymbol{q}\boldsymbol{q}'}$ in the second line to simplify the following descriptions.

We now calculate $\boldsymbol{\chi}^{\boldsymbol{q}\boldsymbol{q}'}(i\omega_n)$ in an approximation which is consistent with the VCA.
There are two types of contributions
\begin{align}
\boldsymbol{\chi}^{\boldsymbol{q}\boldsymbol{q}'}(i\omega_n)
= \delta_{\boldsymbol{q}\boldsymbol{q}'}
\boldsymbol{\chi}_0^{\boldsymbol{q}}(i\omega_n)
+ \boldsymbol{\chi}_\textrm{corr}^{\boldsymbol{q}\boldsymbol{q}'}(i\omega_n).
\end{align}
The diagrammatic representation is shown in Fig.~\ref{d1}(a). 
The first term corresponds to the ``bubble'' diagram, which is explicitly represented as  
\begin{align}
&[\boldsymbol{\chi}^{\boldsymbol{q}}_0(i\omega_n)]_{(\sigma l_1 l_2), (\sigma' l_3 l_4)}
\nonumber \\
&= - \delta_{\sigma\sigma'}
T \sum_{\epsilon_{m}}
\mathcal{G}_{l_3 l_1}^{\sigma}(\boldsymbol{q},i\epsilon_{m})
\mathcal{G}_{l_2 l_4}^{\sigma}(\boldsymbol{q},i\omega_{n}+i\epsilon_{m}),
\end{align}
where $\epsilon_{m}$ is the fermionic Matsubara frequency.
The second contribution $\boldsymbol{\chi}_\textrm{corr}^{\boldsymbol{q}\boldsymbol{q}'}$ describes vertex corrections.
The full (reducible) vertex $\Gamma$, in general, depends on three frequencies and two momenta (the momentum transfer $\boldsymbol{Q}$ is fixed at $\boldsymbol{Q}=0$), namely,
$\Gamma=\boldsymbol{\Gamma}^{\boldsymbol{q}\boldsymbol{q}'}(i\omega_{n};i\epsilon_{m},i\epsilon_{m'})$.
In the VCA, the momentum dependence can be droped as in the local approximation to the self-energy.
Furthermore, we neglect the fermionic frequencies, $\epsilon_{m}$ and $\epsilon_{m'}$, keeping only the bosonic frequency $\omega_n$. This approximation has been adopted in the literature~\citep{gammafunction1,gammafunction2}.
The vertex correction is thus expressed as
\begin{align}
\boldsymbol{\chi}_\textrm{corr}^{\boldsymbol{q}\boldsymbol{q}'}(i\omega_n)
= \boldsymbol{\chi}_0^{\boldsymbol{q}}(i\omega_n) \boldsymbol{\Gamma}(i\omega_n) \boldsymbol{\chi}_0^{\boldsymbol{q}'}(i\omega_n).
\label{eq:19}
\end{align}
%
The vertex part $\boldsymbol{\Gamma}(i\omega_{n})$ is represented in terms of the irreducible vertex $\boldsymbol{I}(i\omega_{n})$ using the Bethe-Salpeter equation
\begin{align}
\boldsymbol{\Gamma}(i\omega_n) = \boldsymbol{I}(i\omega_n)
+ \frac{L}{N} \sum_{\boldsymbol{q}} \boldsymbol{I}(i\omega_n) \boldsymbol{\chi}_0^{\boldsymbol{q}}(i\omega_n) \boldsymbol{\Gamma}(i\omega_n).
\end{align}
Figure~\ref{d1}(b) shows the diagrammatic expression of this equation.
$\boldsymbol{I}(i\omega_n)$ is the two-particle counterpart of the self-energy, and therefore is considered to be local.
We compute $\boldsymbol{I}(i\omega_n)$ in the reference system by solving the Bethe-Salpeter equation defined within the cluster.


\section{Results\label{sec:Results}}

\subsection{Finite $T$ phase diagram}

\begin{figure}[t]\centering
\includegraphics[width=\columnwidth]{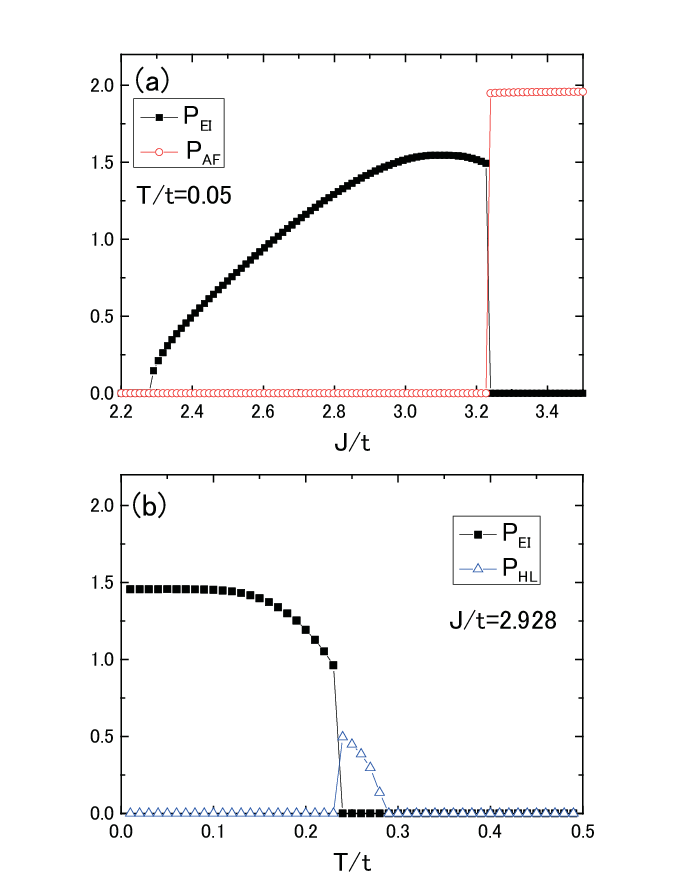}
\caption{(Color online) 
(a) The order parameters as a function of the Hund coupling at $T/t=0.005$, and (b) those as a function of the temperature at $J/t=2.928$. }
\label{op}
\end{figure}

First, we calculate the temperature dependences of the order parameters of the AF, EI and HL states defined by 
$P_{{\rm AF}}=\langle p_{{\rm AF}}\rangle$, $P_{{\rm EI}}=\langle p_{{\rm EI}}\rangle$,
and $P_{{\rm HL}}=\langle p_{{\rm HL}}\rangle$, respectively. 
The order parameters as a function of $J$ are shown in Fig.~\ref{op}(a). 
With increasing $J$, the LS state realized in $J/t \lesssim 2.3$ is changed into 
the EI ordered state through the second-order phase transition. 
This is changed furthermore the HS state associated with the AF order through the first-order phase transition at $J/t\sim3.25$. 
Temperature dependences of the order parameters at  $J/t=2.928$ 
are shown in Fig.~\ref{op}(b). 
With decreasing temperature, a sequential phase
transition occurs; the second-order phase transition from the paramagnetic
phase to the HL ordered phase at $T/t\sim0.28$, and the first-order
transition to the EI phase at $T/t \sim0.24$.

\begin{figure}[t]\centering
\includegraphics[scale=0.55]{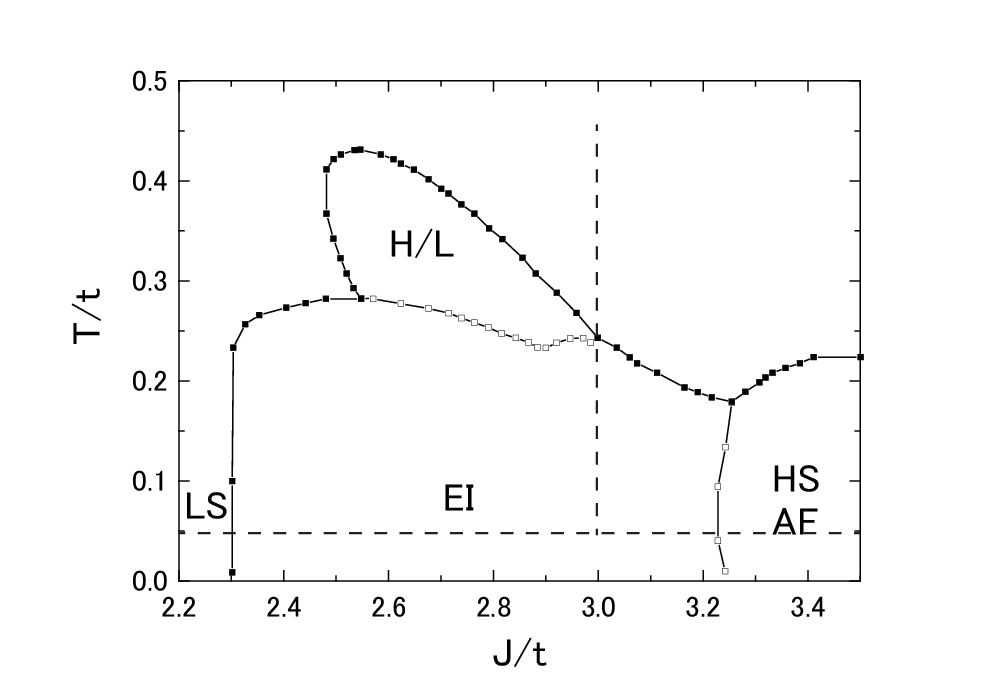}
\caption{
The finite temperature phase diagram. 
Filled and open squares represent the second- and first-order phase transitions,
respectively. 
Horizontal and vertical lines represent the parameter sets adopted in Figs.~\ref{opt1} and \ref{opt2}, respectively. 
}
\label{phasediagram}
\end{figure}

Calculated results are summarized in the finite-temperature phase diagram shown in Fig.~\ref{phasediagram}, in
which open and filled symbols represent the first- and second-order
phase transitions, respectively. 
In the regions of small and large $J/t$, the LS and HS states are realized, respectively, as expected from the electronic structures in a single site. 
Between the two phases, the EI phase appears in low temperatures. 
The HL ordered phase appears above the EI phase around $2.5<J/t<3.0$. 
%
%
In Ref.~\citep{tatsuno2016}, the finite-temperature phase diagram was obtained by applying the mean-field approximation to the effective pseudo-spin model derived from the two-orbital Hubbard model. 
It was shown that the HL phase is realized down to $T=0$ in contrast to Fig.~\ref{phasediagram}.  
This difference is attributed to the fact that the spin entropy in the HL phase is overestimated in the mean-field approximation in Ref.~\citep{tatsuno2016}.

\subsection{One particle excitation spectra}

\begin{figure}\centering
\includegraphics[scale=0.8]{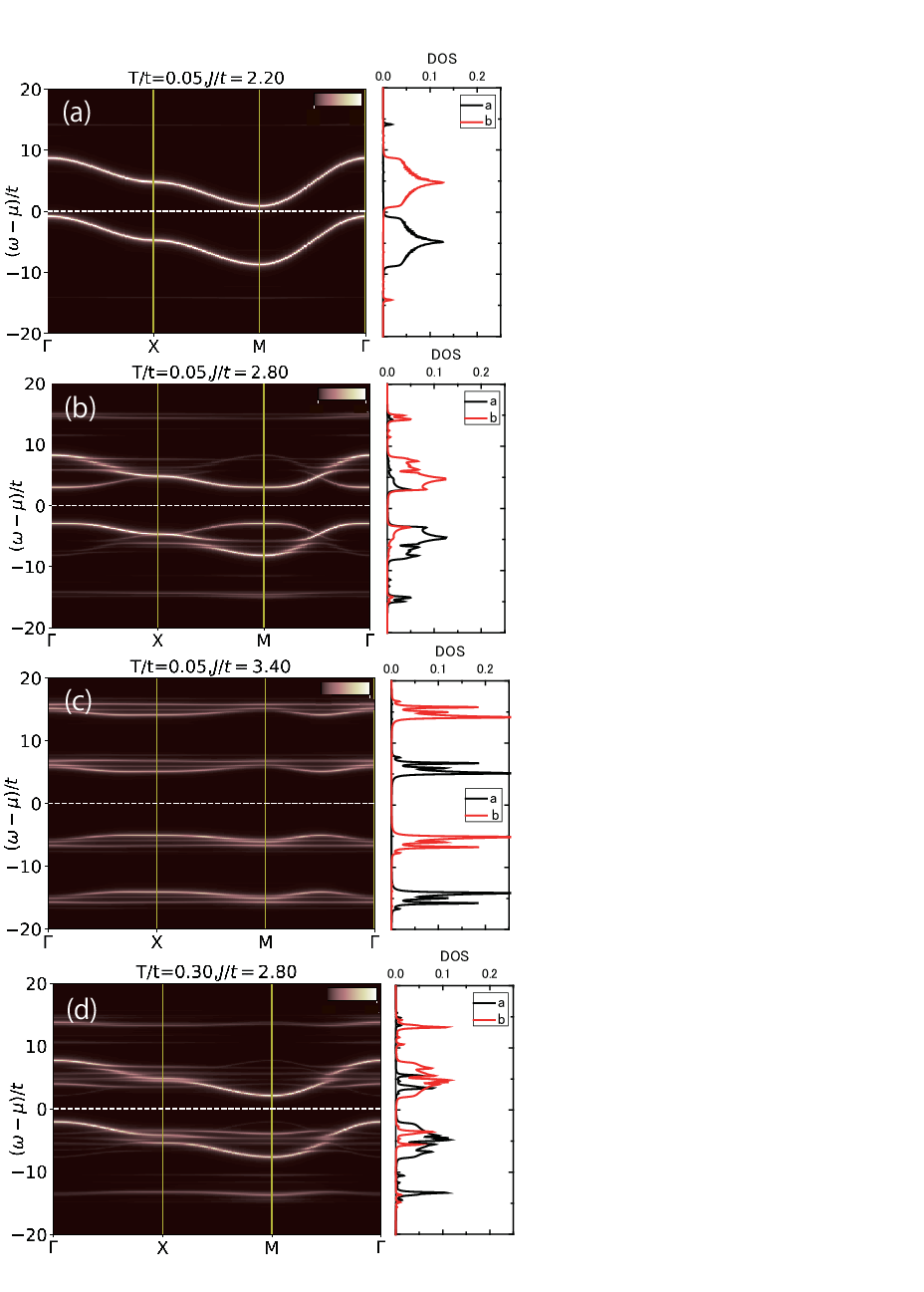}
\caption{(Color online) Intensity plots of the one-particle excitation spectra
in (a) $J/t=2.2$, (b) $2.8$, (c) $3.4$ and (d) $2.8$. 
We set $T/t=0.05$ in (a)-(c) and $0.3$ in (d), and $\eta=0.1$. 
Right panel in each figure shows
DOS where black and red lines represent the $a$ and $b$ orbital
components, respectively. }
\label{dos}
\end{figure}

We calculate the single-particle excitation spectra defined by  
\begin{align}
\rho^{\sigma}(\boldsymbol{k},\omega)
=-\frac{1}{\pi} \textrm{Im} \sum_{\alpha} g_{\alpha}^{\sigma}(\boldsymbol{k},\omega+i\eta),
\end{align}
where $g_{\alpha}^{\sigma}(\boldsymbol{k},\omega)$ is the lattice Green's function
in the original Brillouin zone
for the orbital $\alpha(=a,b)$. 
Following the CPT~\citep{CPT2000}, we compute $g_{\alpha}^{\sigma}(\boldsymbol{k},\omega)$ from the cluster Green's function $\mathcal{G}_{ll'}^{\sigma}(\boldsymbol{k},\omega)$
by 
\begin{align}
g_{\alpha}^{\sigma}(\boldsymbol{k},\omega) & =\frac{1}{L/2}\sum_{l,l' \in \alpha}e^{-i\boldsymbol{k}\cdot(\boldsymbol{r}_{l}-\boldsymbol{r}_{l'})}\mathcal{G}_{ll'}^{\sigma}(\boldsymbol{k},\omega),
\end{align}
where the summation for $l \equiv(i,\alpha)$ and $l' \equiv(j,\alpha)$ are taken for all sites in the cluster, keeping the orbital index $\alpha$.
The intensity maps of the spectral functions and the density of states (DOS) at $T/t=0.05$ are shown in Fig. \ref{dos}. 
The ground state for the results in Figs.~\ref{dos}(a), (b) and (c) are the LS, EI and HS states, respectively. 
In the LS state shown in Fig.~\ref{dos}(a), the valence and conduction bands mainly consist of the $a$ and $b$ orbitals, respectively, and the Fermi level is located inside of the band gap. 
Small component of the $b$ ($a$) orbital in the valence (conduction)
band is induced by the pair-hopping interaction. 
The dispersion relations are almost reproduced by the non-interacting tight-binding model, i.e. the system is a band insulating state. 
In the EI state shown in Fig.~\ref{dos}(b), 
the hybridization between the $a$ and $b$ orbitals are realized. 
The one-particle excitation spectra mainly consist of the two parts: 
the cosine-like bands similar to the ones in the LS phase [see Fig.~\ref{dos}(a)], and the new bands which appear around M ($\Gamma$) points in the valence (conduction) band.  
The latter is termed the shadow bands from now on. 
The bottom of the conduction band and the top of the valence band are flatten in comparison with the simple cosine-bands in the LS state. 
As a result, the energy gap is enlarged. 
These characteristics originate from the staggered EI order which corresponds to the band mixing between the top of the valence band around the $\Gamma$ point and the bottom of the valence band round the M point. 
In the HS state shown in Fig.~\ref{dos}(c), 
 the hybridization between the two orbitals does not appear. 
The each orbital band is separated into the upper-Hubbard band (UHB) and the lower-Hubbard band (LHB), 
and the Fermi Level is located between the LHB for the $a$ orbital and the UHB for the $b$ orbital. 
That is, the system is identified as a Mott insulating state. 

The results in the higher-temperature HL state are shown in  Fig.~\ref{dos}(d). 
The $a$ $(b)$-orbital component in the conduction (valence) band is not due to the hybridization effect but mainly due to 
the HL order and the thermal effect. 
When we compare the results in Fig.~\ref{dos}(d) with Fig.~\ref{dos}(b), the shadow bands appear only in the EI state. 
On the other hand, the band structures and DOS in Fig.~\ref{dos}(d) are almost explained by the simple averages of the results in the LS state and the HS state shown in Figs.~\ref{dos}(a) and (c), respectively. 


\subsection{Optical conductivity spectra}

\begin{figure}\centering
\includegraphics[scale=0.5]{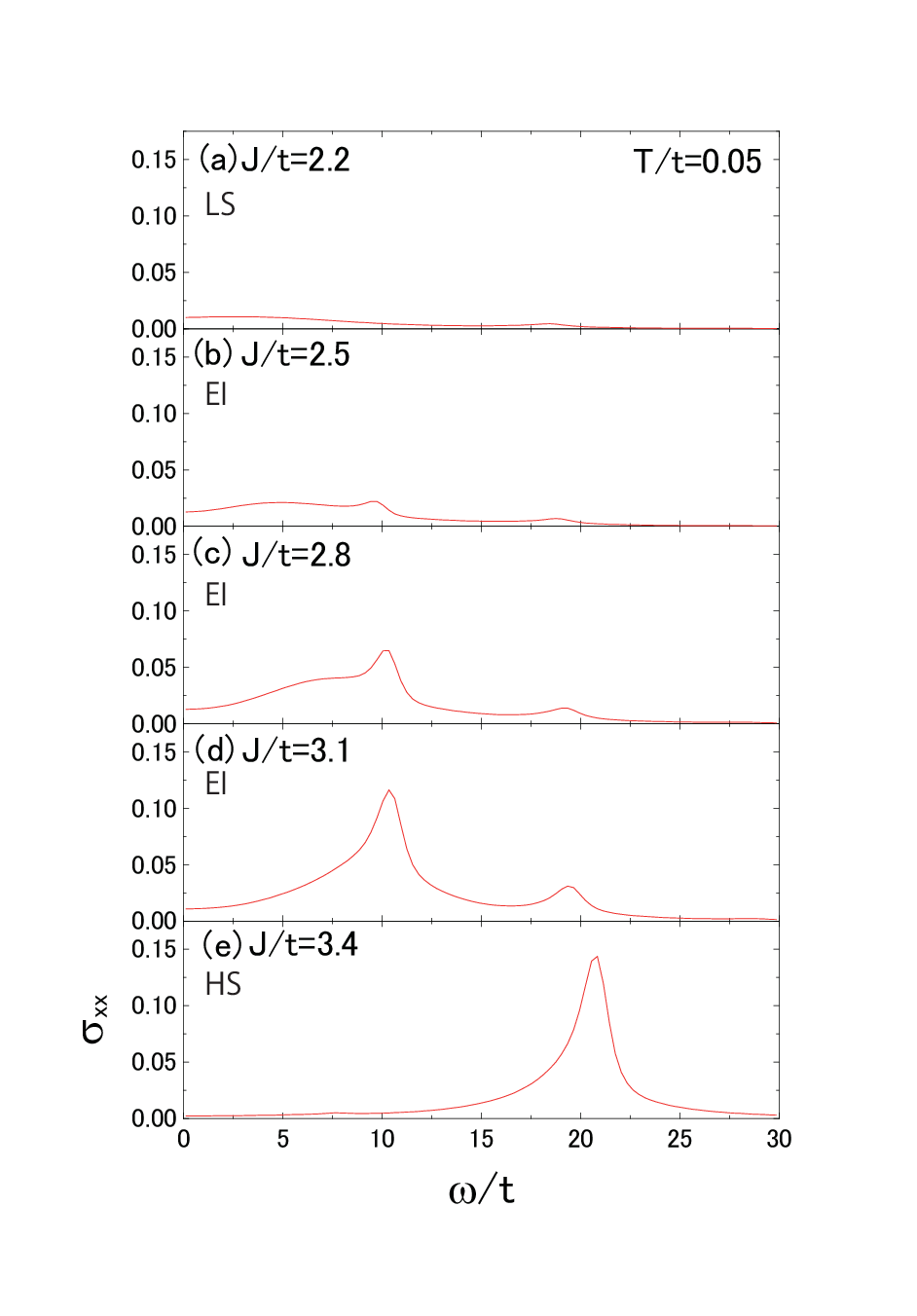}
\caption{(Color online) Optical conductivity spectra for several values of
$J$ at $T/t=0.05$. 
}
\label{opt1}
\end{figure}


The optical conductivity spectra for several values of $J/t$ at $T/t=0.05$
are shown in Fig.~\ref{opt1}. 
In the LS phase, there are no remarkable peaks in $\sigma(\omega)$, and the total spectral weight is small. 
This is attributed to the fact that the valence band consists mainly of the $b$ orbital, and there is no electron
hopping between the nearest neighboring $a$ and $b$ orbitals. 
Weak spectra are caused by the thermal fluctuation and the pair-hopping interaction. 
%
In the EI phase, a new peak around $\omega/t\sim10$ and continuum around $3 \lesssim \omega/t \lesssim 8  $ appear. 
Since these are related to the EI order as explained below, we term these the EI components from now on. 
The intensity of the EI component increases with increasing $J/t$ in the EI phase. 
Small peak around $\omega/t=20$ exists in the EI phase, and changes into a sharp peak in the HS phase, in which the EI component almost disappears. 
This peak is termed the Hubbard component from now on. 

\begin{figure}\centering
\includegraphics[scale=0.4]{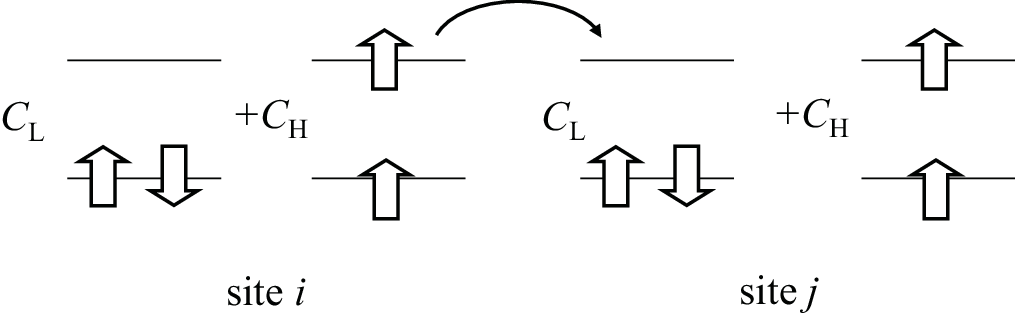}
\caption{
Schematic optical excitation processes in the EI component. 
Vertical bars and arrows represent the orbitals and the spin directions, respectively. 
Symbols $C_{\rm L}$ and $C_{\rm H}$ are the coefficients of the wavefunction in the EI state. (see text).  
}
\label{process}
\end{figure}
These EI components are attributed to the mixing of the $a$ and $b$ orbitals. 
Since the wave function in the EI phase is represented by the linear combinations 
of the LS and HS states, e.g. $C_{\rm L}|a^0 b^2 \rangle+C_{\rm H}|a^1 b^1 \rangle$ with coefficients $C_{\rm L}$ and $C_{\rm H}$ (see Fig.~\ref{process}(a)), 
new optical-excitation processes in the nearest neighboring sites open. 
As explained in the previous section, the one-particle excitation spectra shown in Fig.~\ref{dos}(b) mainly consist of the cosine-like bands and the shadow bands. 
The sharp EI components around $\omega/t \sim 10$ originates from the transitions between the cosine--type conduction and valence bands. This is prohibited in the LS phase, since the $a$ ($b$) orbitals are almost occupied (unoccupied), and is permitted in the EI phase. 
Since the two bands almost overlap by the parallel shift of $D$, the optical excitations between the bands induce the sharp peak structure around $\omega=D$. 
On the other hand, the continuum in the EI component is attributed to the optical transitions between the cosine-like valence (conduction) band and the shadow conduction (valence) band around the point $\Gamma$ (M). 
These excitations spread from the band gap, which is about $3t$, up to about $10t$. 
The spectral weight in the EI component is approximately estimated to be $|C_{\rm L} C_{\rm H}|^2$. 
The Hubbard component is identified as the intersite electronic excitation 
between the nearest neighboring HS states with the antiferromagnetic spin configuration, 
and the excitation energy is of the order of $U$. 
This excitation is possible even in the EI phase, and the spectral weight is approximately given by $|C_{\rm H}|^4$. 

\begin{figure}[t] 
\includegraphics[scale=0.5]{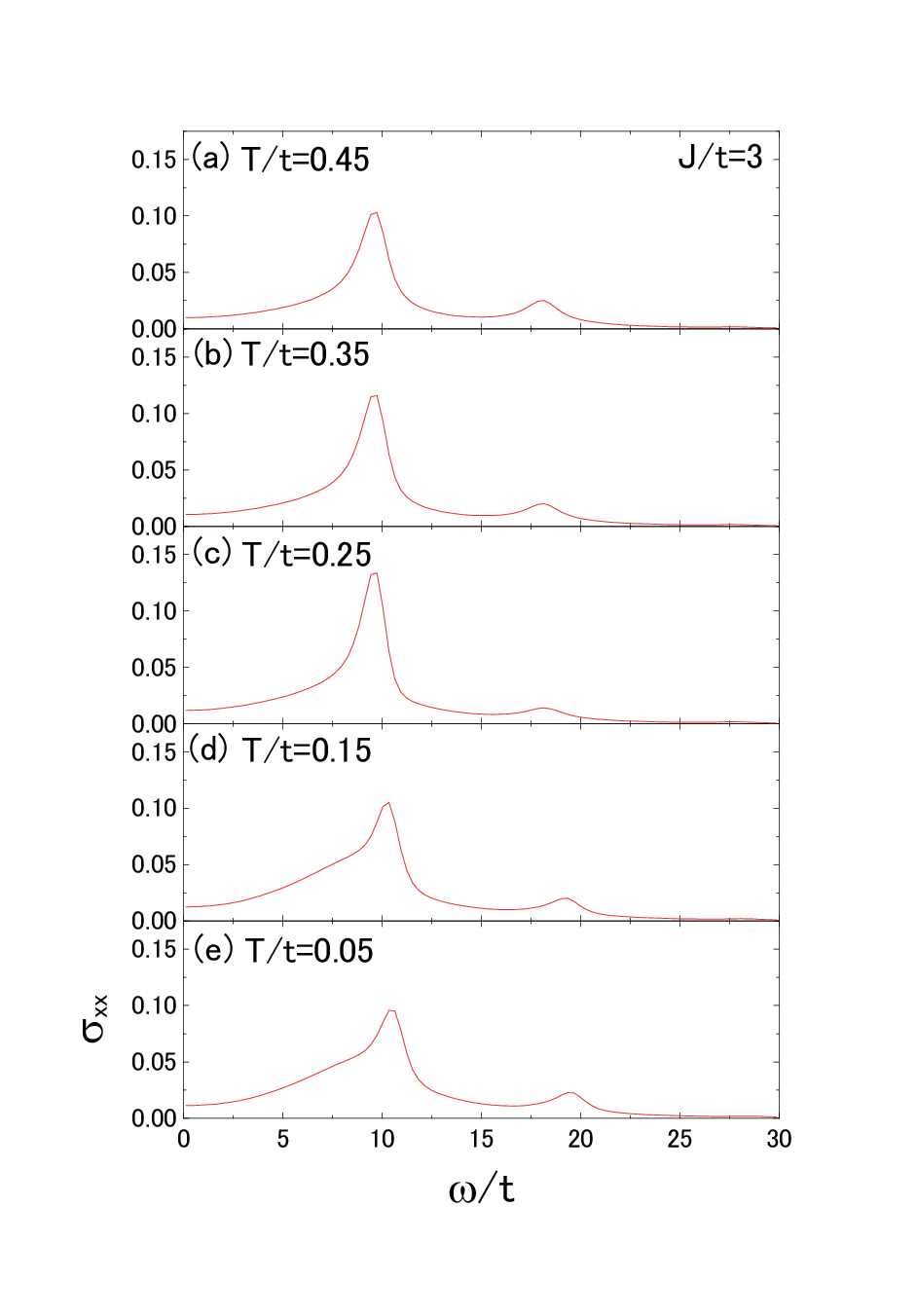}
\caption{(Color online) Optical conductivity spectra for several values of
$T$ at $J/t=3.00$. 
}
\label{opt2}
\end{figure}
%
In Fig.~\ref{opt2}, we show the temperature dependence of the optical
conductivity spectra at $J/t=3$, in which the EI state is the ground state and the EI-ordering temperature is $T/t=0.25$. 
A whole feature of the optical spectra do not show remarkable changes; 
the EI component and the Hubbard component exist in all temperature range. 
With increasing temperature from $T/t=0.05$,
the sharp peak around $\omega/t=10$ becomes remarkable, and the continuum in $3 \lesssim \omega/t \lesssim 8$ is reduced. 
These behaviors correspond to the  one-particle excitation spectra shown in Fig.~\ref{dos}(d), where 
although the two-orbital components are mixed, the shadow bands in the occupied (unoccupied) band at M ($\Gamma$) points shown in Fig.~\ref{opt1}(b) do not appear. 
This mixing between the two-orbital components originate from the thermal effect through the Boltzman factor, i.e. the local electronic states with the LS and HS configurations are mixed at each site, and the optical excitation between the NN states are realized, when the LS and HS states are located in the neighboring sites. 
The intensity of the EI component is of the order of $(e^{-\beta \Delta E})^2$ where $\Delta E$ is the energy difference between the local LS and HS configurations.

\begin{figure}
\includegraphics[scale=0.4]{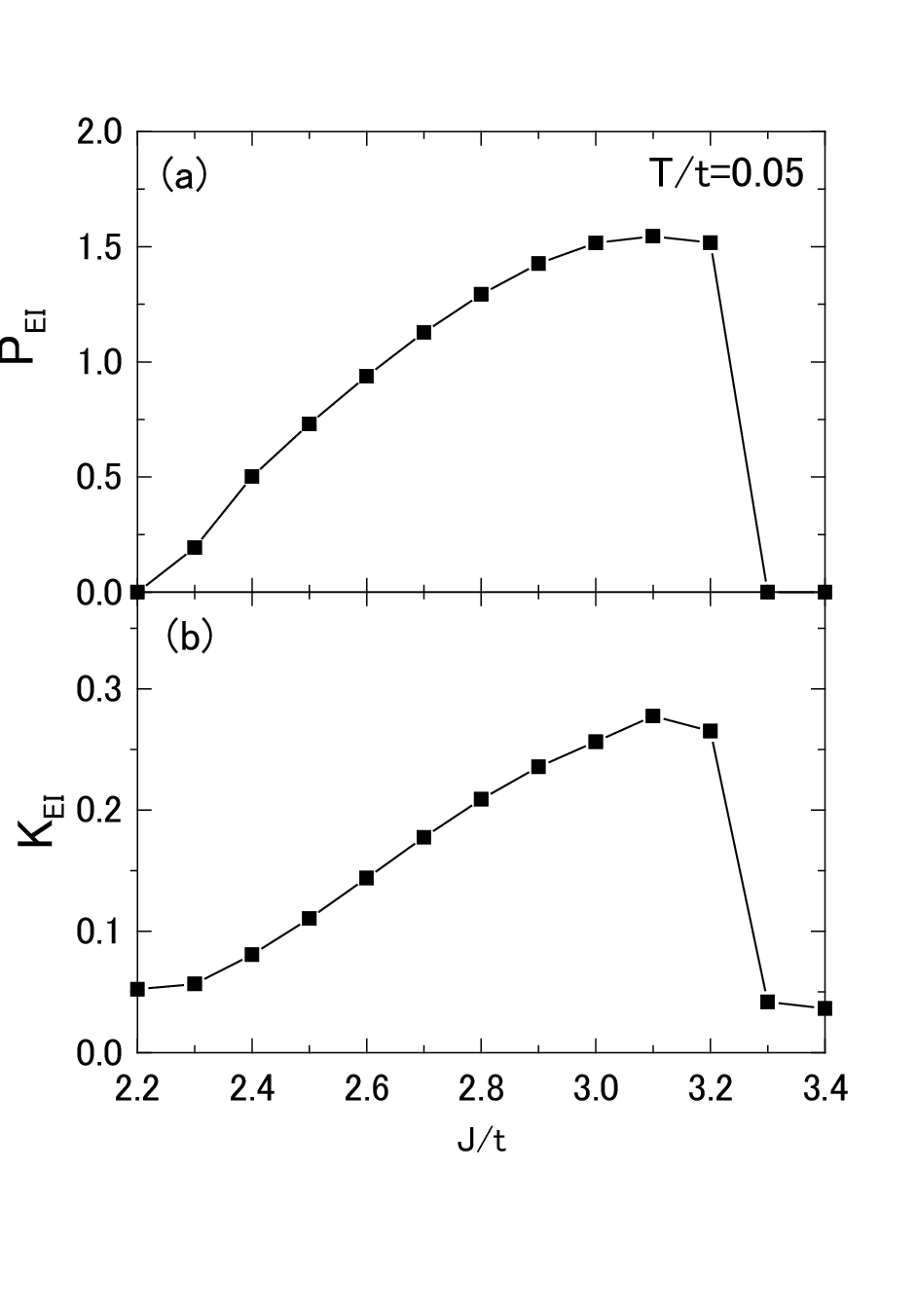}
\caption{(a) The EI order parameter and (b) integrated intensity of the EI component in the optical conductivity spectra as functions of $J/t$ at $T/t=0.05$.}
\label{opt3}
\end{figure}

We analyze qualitatively the EI component in the optical conductivity spectra. 
The integrated intensities of the EI peak defined by $K_{\rm EI}=\int_{0}^{\omega_{c}}\sigma_{xx}(\omega)d\omega$ 
are calculated. 
The cutoff frequency is set to $\omega_{c}/t=15$ which is located between the EI and Hubbard components. 
We checked that the results are not changed much, when $\omega_{c}/t$ is changed between $12$ and $18$. 
The integrated intensities at $T/t=0.05$ are plotted as a function of $J$ in Fig.~\ref{opt3}, where the EI
order parameters are also presented.  
There is a good correspondence between the two results. 
Finite values of $K_{\rm EI}$ at $J/t=2.2$ and $3.4$ are attributed to $\eta$ for the analytical continuation. 
The present analyses imply that the EI peak intensity reflects the EI order parameters in
low $T$ region, and is available to identify the EI state. 


\section{Concluding remarks}

In summary, we study the optical responses in the EI states. 
The VCA method is applied to the  two-orbital Hubbard model with a finite energy difference  
between the orbitals. 
In the analyses of the ground state, we consider possibilities of the LS state, the HS-AF state, the EI state and the H/L state. 
The optical conductivity spectra are formulated by the Green's function method, where the vertex correction is taken into account. 
We find the characteristic components in the optical spectra attributed to the quantum hybridization of the conduction and valence bands. There is a good correspondence between the EI order parameter and the integrated weight of this component in low temperatures. 
We propose that this component is available to identify the EI state in the real systems.

\begin{acknowledgments}
This work was supported by JSPS KAKENHI Grant No. 15H02100, No. 17H02916,
No. 18H05208, No. 18H01158, and No. 18H04301 (J-Physics). The computation
in this work has been done using the facilities of the Supercomputer
Center, the Institute for Solid State Physics, the University of Tokyo.
\end{acknowledgments}





\end{document}